\documentclass{article}

\def\xxinput#1{\input#1}

\xxinput{vsolj02.sty}

\usepackage[dvipdfmx]{graphicx}
\usepackage[comma,colon]{natbib}

\usepackage[OT2,T1]{fontenc}

\def\cite{\citealt}
\setcitestyle{aysep={}}

\newcounter{author}
\def\altaffilmark#1{$^{#1}$}
\def\altaffiltext#1{$^{#1}$\,}

\def\authorcount#1#2{{\refstepcounter{author}\label{#1}
                     \altaffiltext{\ref{#1}}{#2}}}

\begin{document}

\begin{center}

\title{SDSS J094002.56$+$274942.0: an SU UMa star with an orbital period of}
\vskip -2mm
\title{3.92 hours and an apparently unevolved secondary}

\author{
        Taichi~Kato\altaffilmark{\ref{affil:Kyoto}},
        Tonny~Vanmunster\altaffilmark{\ref{affil:Vanmunster}}
}

\authorcount{affil:Kyoto}{
     Department of Astronomy, Kyoto University, Sakyo-ku,
     Kyoto 606-8502, Japan \\
     \textit{tkato@kusastro.kyoto-u.ac.jp}
}

\authorcount{affil:Vanmunster}{
     Center for Backyard Astrophysics Belgium, Walhostraat 1A,
     B-3401 Landen, Belgium \\
     \textit{tonny.vanmunster@gmail.com}
}

\end{center}

\begin{abstract}
\xxinput{abst.inc}
\end{abstract}

\section{Introduction}\label{sec:intro}

   SU UMa-type dwarf novae are a class of cataclysmic variables
(CVs) which show superhumps during long, bright outbursts
(superoutbursts) [For general information of CVs and subclasses,
see e.g., \citet{war95book}].
This phenomenon is widely believed to be
the consequence of the 3:1 resonance
\citep{whi88tidal,hir90SHexcess,lub91SHa} when the disk expands
to this radius during a superoutburst.  Whether the 3:1 resonance
is the cause or the consequence of a superoutburst has been
discussed, and a series of analyses of the Kepler data
of V1504 Cyg and V344 Lyr \citep{osa13v1504cygKepler,
osa13v344lyrv1504cyg,osa14v1504cygv344lyrpaper3} established
that the 3:1 resonance is the cause of a superoutburst
as originally proposed by \citet{osa89suuma}.

   The size of the accretion disk is limited by tidal truncation,
or in extreme cases, could be by the size of the Roche lobe.
The size of the Roche lobe is a function of the mass ratio
$q=M_1/M_2$ and it has long been discussed, both observationally
and theoretically, what is the upper limit of $q$ that enables
the 3:1 resonance.  Numerical simulations such as by
smoothed particle hydrodynamics (SPH) gave a limit of
$q<$0.25--0.33 (e.g. \cite{whi91SH,mur98SH,mur00SHintermediateq,smi07SH}).
CVs with longer orbital periods ($P_{\rm orb}$) have larger $q$
and there is a natural upper limit for $P_{\rm orb}$ for
objects showing superhumps.  The limit $q$=0.25 generally
corresponds to the period gap in CVs.
This limit originally appeared to fit SU UMa-type dwarf novae
very well, among which systems above the period gap
were very exceptional.
It has been well-known, however, superhumps are present
in longer-$P_{\rm orb}$ systems in novalike variables (systems
with a thermally stable high-state disk), such as in
\citet{bru22CVTESS1,ste22bgtri,bru23CVTESS2} using
modern Transiting Exoplanet Survey Satellite
(TESS: \cite{ric15TESS}) data.
Although the case of RZ Gru ($P_{\rm orb}$=0.4563~d)
in \citet{bru22CVTESS1} may be exceptional and it may have
shown a phenomenon different from (ordinary) superhumps,
many novalike variables are known to show superhumps
at least up to $P_{\rm orb}$=0.18~d \citep{bru23CVTESS2}.
Since the majority of long-$P_{\rm orb}$
systems showing superhumps are novalike systems, there has
been an idea that a weaker effect of a resonance requires
longer time to develop, and systems in high states for
a sufficiently long time (i.e. novalike variables) only can
show superhumps.  In this interpretation, the high states
(outbursts) in dwarf novae are not long enough to develop
superhumps, somewhat in line with the above-mentioned idea
that the 3:1 resonance is the consequence of a long outburst,
not the cause.

   In recent years, however, a number of SU UMa-type dwarf novae
have been discovered above the period gap.
The only traditionally known system was TU Men \citep{sto84tumen}\footnote{
   The case of U Gem was claimed by \citet{sma04ugemSH}, but
   T.K. considered it questionable \citep{Pdot5} and
   the case is far from being established
   [see also \citet{pat05SH}].
}.
The $q$ value for TU Men, however, is unfortunately
not well-determined \citep{sma06tumen}.  There is evidence for
CNO-processing in this object \citep{god21ssaurtumen} and
TU Men may not be considered an ordinary CV following
the standard evolutionary track.
Modern examples without evidence of an evolved secondary include
CRTS J035905.9$+$175034 [$q$=0.281(15): \citet{lit18j0359}] and
BO~Cet [$q$=0.31--0.34: \citet{kat21bocet,kat23bocet}].
Superhumps in both objects started to develop soon after
superoutbursts reached the plateau phase
as in other SU UMa stars \citep{osa13v1504cygKepler}.
There was no indication of a significant delay in the development
of superhumps, which would be expected if the 3:1 resonance is
the consequence of a long outburst.

   SDSS J094002.56$+$274942.0 was spectroscopically selected
as a CV \citep{szk07SDSSCV6}.  The spectrum by \citet{szk07SDSSCV6}
was that of an ordinary dwarf nova.
\citet{kra10j0940} noticed an outburst detected by
Catalina Real-Time Transient Survey (CRTS) in 2009 and
performed time-resolved photometry.  \citet{kra10j0940} detected
orbital modulations (ellipsoidal variations) with a period of
0.16352~d during the fading part of this outburst.
Although \citet{kra10j0940} suggested that the disk or
the hot spot may have been (partially) eclipsed during the brighter
stages of this outburst, the nature of the observed dip was unclear
due to the poor phase coverage.  \citet{kra10j0940} discussed
that TiO bands expected from the secondary type (M4--5) were
invisible in the spectrum of \citet{szk07SDSSCV6}.
\citet{kra10j0940} noted a possibility of an early M-type
secondary, which would suggest an evolved core,
with little TiO absorption.
Based on this identification of $P_{\rm orb}$,
SDSS J094002.56$+$274942.0 has long been regarded as
an SS Cyg-type dwarf nova.
\citet{hou20LAMOSTCV1} observed this object three times
by the LAMOST survey.  No C\textsc{iii}/N\textsc{iii} emission
lines were recorded.  Fe\textsc{ii} emission lines were recorded
once and H$\alpha$ lines was double-peaked on a single occasion.
All spectra were obtained in quiescence and one of them
was presented in \citet{han20v367pegv537peg}.

   During an inspection of Zwicky Transient Facility
(ZTF: \cite{ZTF})\footnote{
   The ZTF data can be obtained from IRSA
$<$https://irsa.ipac.caltech.edu/Missions/ztf.html$>$
using the interface
$<$https://irsa.ipac.caltech.edu/docs/program\_interface/ztf\_api.html$>$
or using a wrapper of the above IRSA API
$<$https://github.com/MickaelRigault/ztfquery$>$.
} data, one of the authors (T.K.) noted that ZTF incidentally
obtained time-resolved photometry during a long outburst
in 2019 February.  Combined with time-resolved observations
by the second author (T.V.) reported to VSNET Collaboration
\citep{VSNET}, T.K. found that this outburst was a superoutburst
and that this object showed superhumps (vsnet-chat 9373)\footnote{
   $<$http://ooruri.kusastro.kyoto-u.ac.jp/mailarchive/vsnet-chat/9373$>$.
}.

\section{Data analysis}

   The observations by T.V. were obtained with a 0.40-m f/5.1
Newton telescope and an unfiltered Starlight Xpress Trius SX-46 CCD camera
with KAF-16200 (3$\times$3 bin) chip, located in Extremadura, Spain.
We used ZTF and Asteroid Terrestrial-impact
Last Alert System (ATLAS: \cite{ATLAS}) forced photometry
\citep{shi21ALTASforced} data for our analysis.
Some unfiltered snapshot observations reported to VSOLJ
were also used (hereafter CCD).
Some ATLAS and CCD observations had false bright detections,
which were removed after comparison with the observations
on the same night or with other observers.
Although TESS also observed this object in quiescence,
these observations were strongly contaminated by nearby brighter
stars and we did not use them since they were not particularly
useful in determining $P_{\rm orb}$ in the presence of
ZTF and ATF observations with a longer baseline.
The log of time-resolved observations is listed in table \ref{tab:log}.
When analyzing the quiescent data, we used locally-weighted polynomial
regression (LOWESS: \cite{LOWESS}) to remove long-term trends.
The periods were determined using the phase dispersion
minimization (PDM: \cite{PDM}) method, whose errors were
estimated by the methods of \citet{fer89error,Pdot2}.

\xxinput{obs.inc}

\section{Results}

\subsection{Long-term behavior}

\begin{figure*}
\begin{center}
\includegraphics[width=16cm]{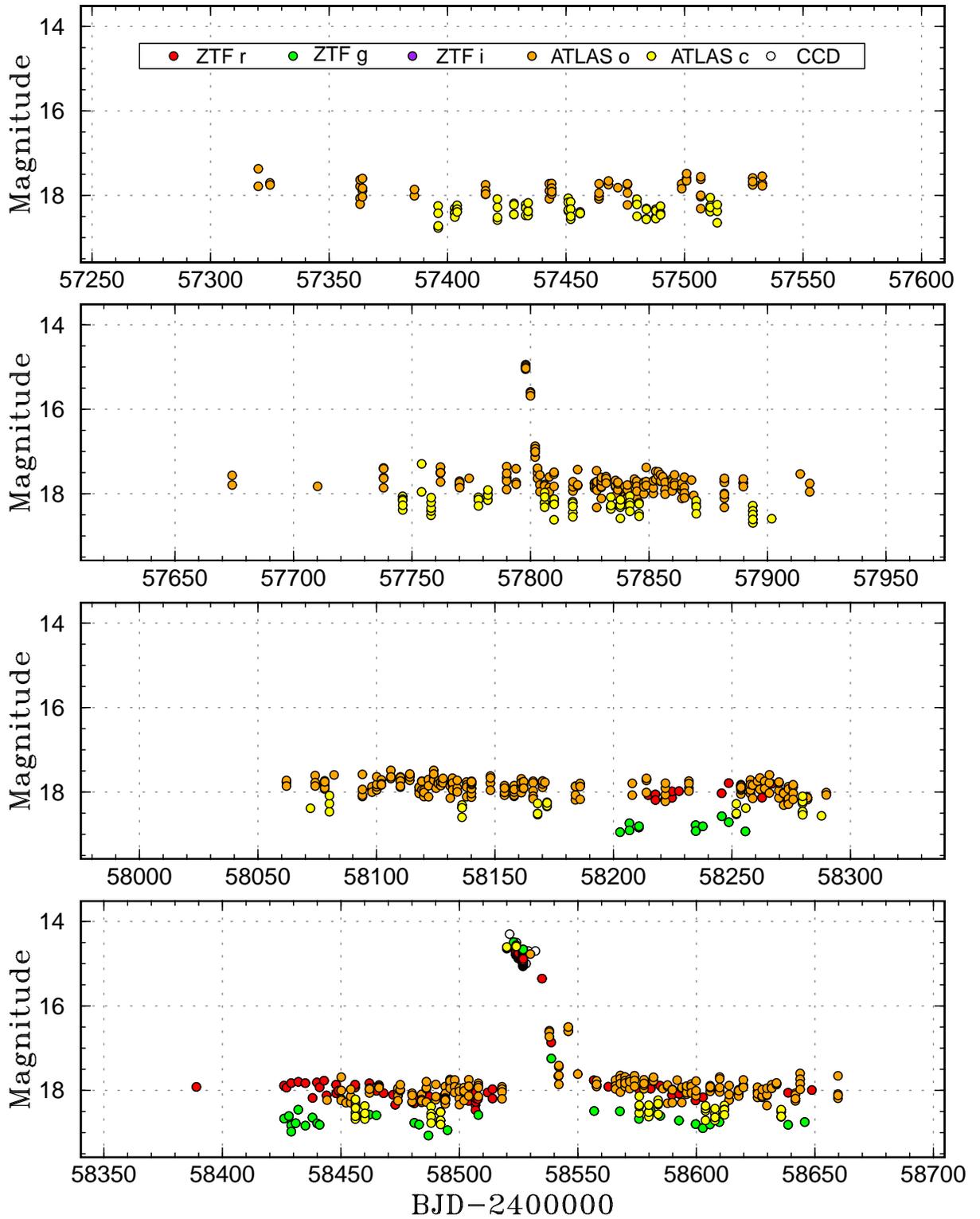}
\caption{
   Light curve of SDSS J094002.56$+$274942.0 in 2015--2019.
CCD refers to unfiltered snapshot observations.
}
\label{fig:lc1}
\end{center}
\end{figure*}

\begin{figure*}
\begin{center}
\includegraphics[width=16cm]{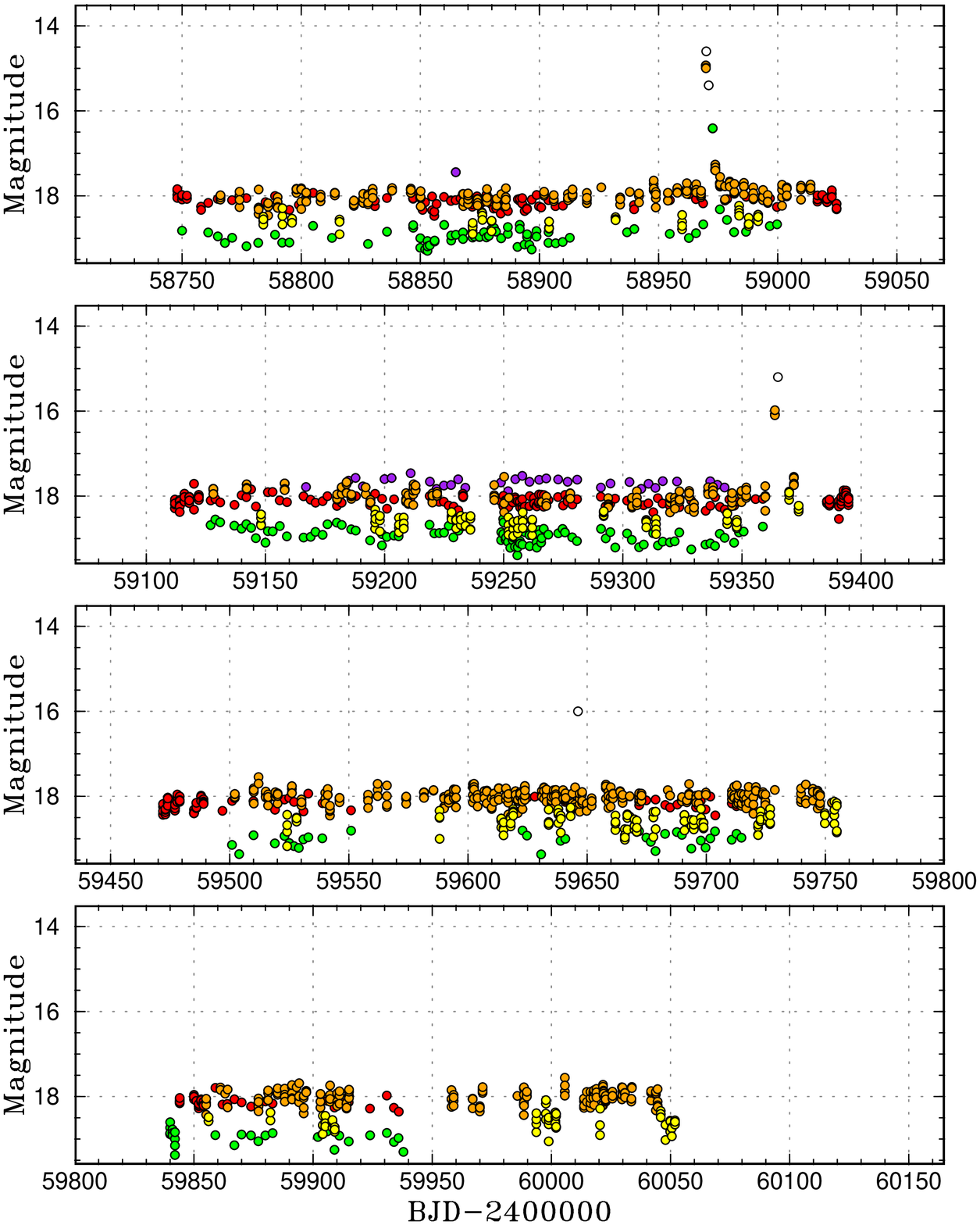}
\caption{
   Light curve of SDSS J094002.56$+$274942.0 in 2019--2023.
The symbols are the same as in figure \ref{fig:lc1}.
}
\label{fig:lc2}
\end{center}
\end{figure*}

   Long-term light curves are shown in figures \ref{fig:lc1}
and \ref{fig:lc2}.  This object showed outbursts relatively
infrequently.  The 2019 February outburst (fourth panel
in figure \ref{fig:lc1}) was a long one, which later turned
out to be a superoutburst (subsection \ref{sec:SH}),
while the other three were short.
The 2009 outburst observed by \citet{kra10j0940} was also
a short one.  The rarity of outbursts suggests
a low mass-transfer rate.

\begin{figure*}
\begin{center}
\includegraphics[width=16cm]{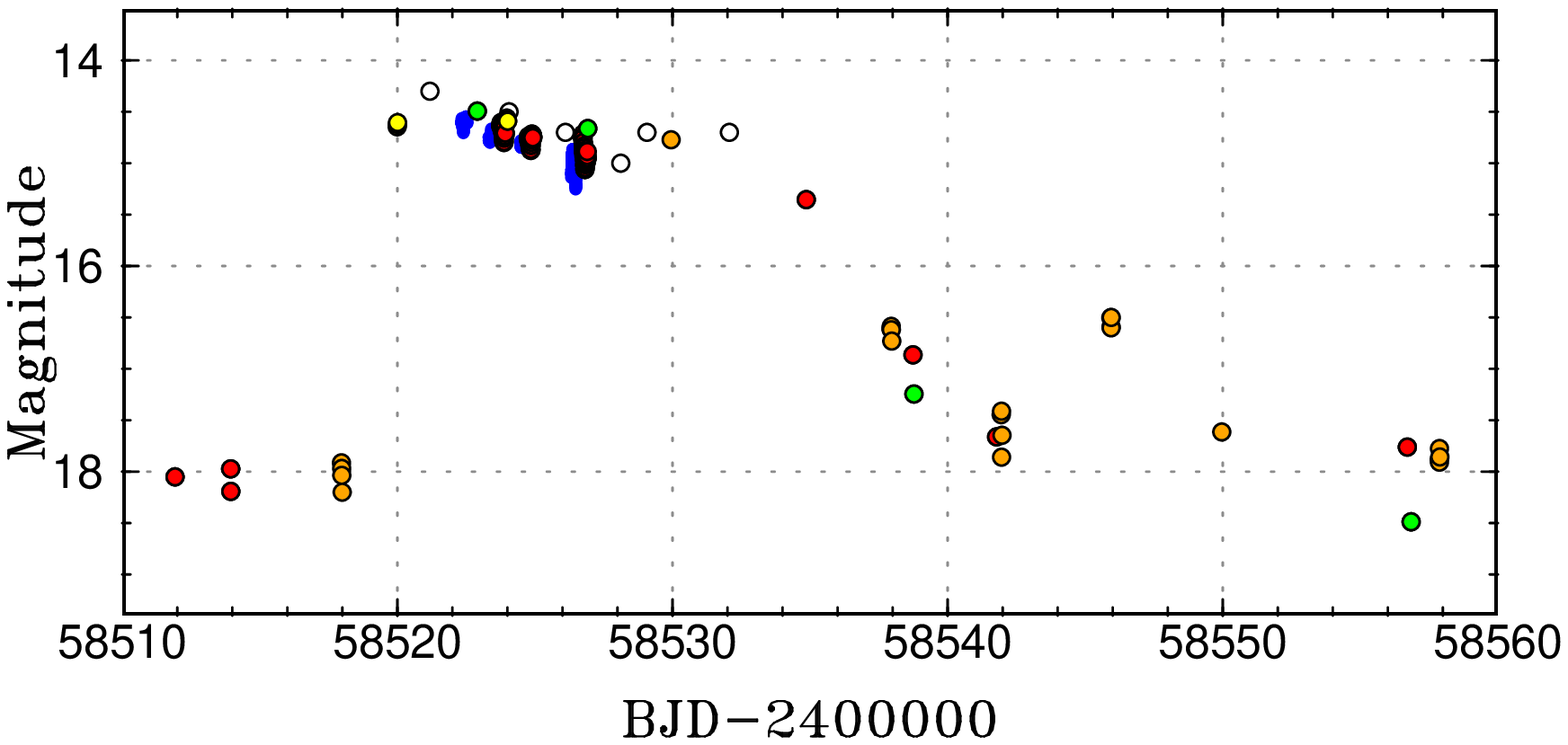}
\caption{
   Light curve of SDSS J094002.56$+$274942.0 during
the 2019 February superoutburst.  Blue plots are time-resolved
photometry by T.V.
The other symbols are the same as in figure \ref{fig:lc1}.
}
\label{fig:lcso}
\end{center}
\end{figure*}

   The details of the 2019 February long outburst are shown in
figure \ref{fig:lcso}.  This outburst had an approximate
duration of 20~d and was followed by a rebrightening
on 2019 March 3 (BJD 2458546).  The overall shape of
this outburst is compatible with that of a superoutburst,
which will be examined in subsection \ref{sec:SH}.

\subsection{Orbital period and profile}

\begin{figure*}
\begin{center}
\includegraphics[width=14cm]{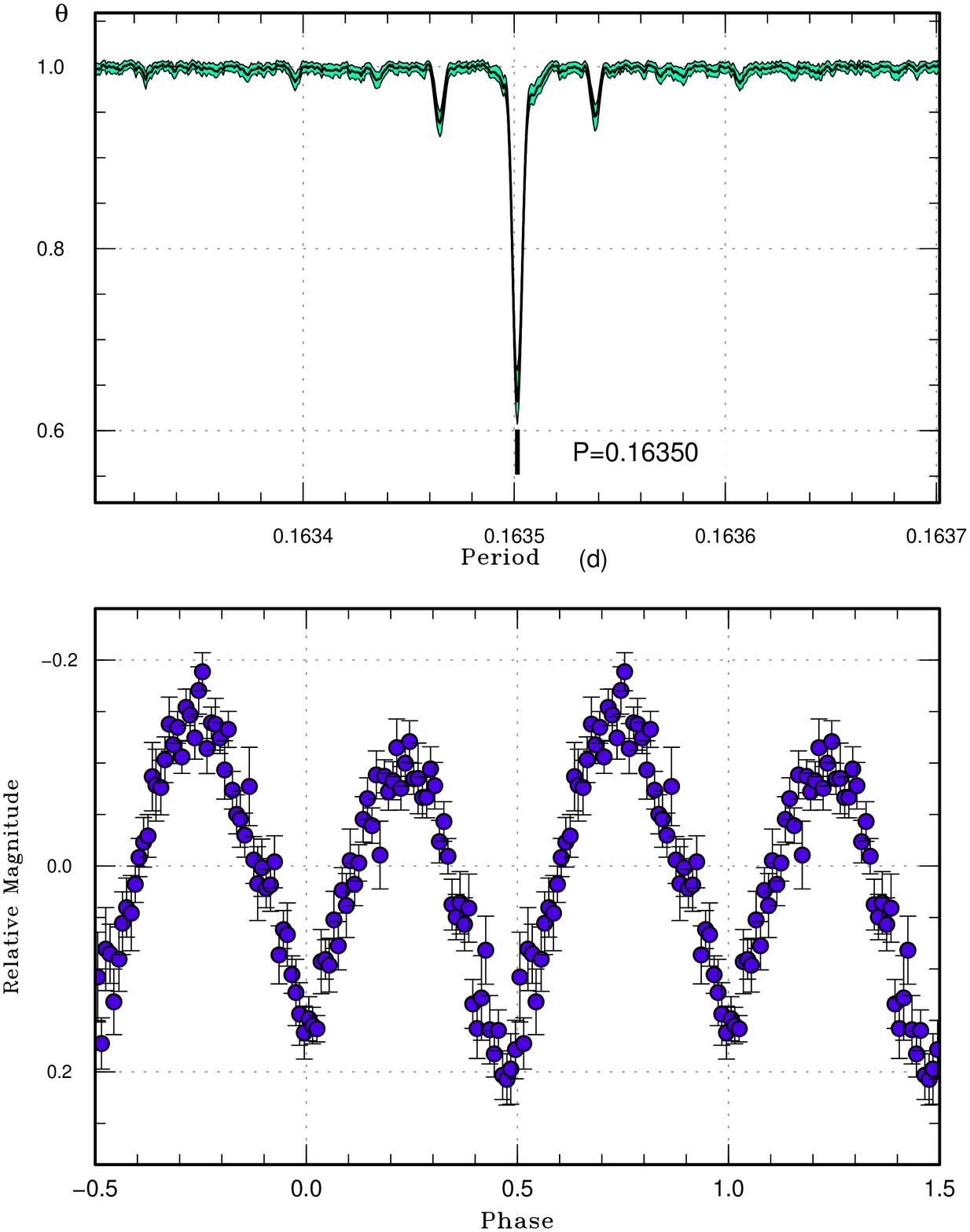}
\caption{
   Orbital profile of SDSS J094002.56$+$274942.0 in quiescence.
   (Upper): PDM analysis.  The bootstrap result using
   randomly contain 50\% of observations is shown as
   a form of 90\% confidence intervals in the resultant 
   $\theta$ statistics.
   (Lower): Phase plot.
}
\label{fig:porbpdm}
\end{center}
\end{figure*}

   We used ZTF and ATLAS observations in quiescence
(all bands were combined after zero-point adjustments)
and obtained the same ellipsoidal variations detected by
\citet{kra10j0940} (figure \ref{fig:porbpdm}).
The orbital phase was defined by
\begin{equation}
\mathrm{Min(BJD)} = 2458953.7169(10) + 0.1635015(1) E.
\label{equ:ecl}
\end{equation}
The zero phase was determined by an MCMC analysis
\citep{Pdot2} of the eclipses detected during the 2019
superoutburst (see subsection \ref{sec:SH}) since
eclipses during an outburst are a better indicator of
the center of the disk particularly when a hot spot is present. 
This ephemeris, however, very well expressed the primary
(but shallower) minimum of the ellipsoidal variations
in quiescence.  The phase 0.75 peak was brighter than the phase
0.25 one probably due to the hot spot.  The secondary
(but deeper) minimum occurred somewhat before the phase 0.5.

\subsection{Eclipses and superhumps in superoutburst}\label{sec:SH}

\begin{figure*}
\begin{center}
\includegraphics[width=14cm]{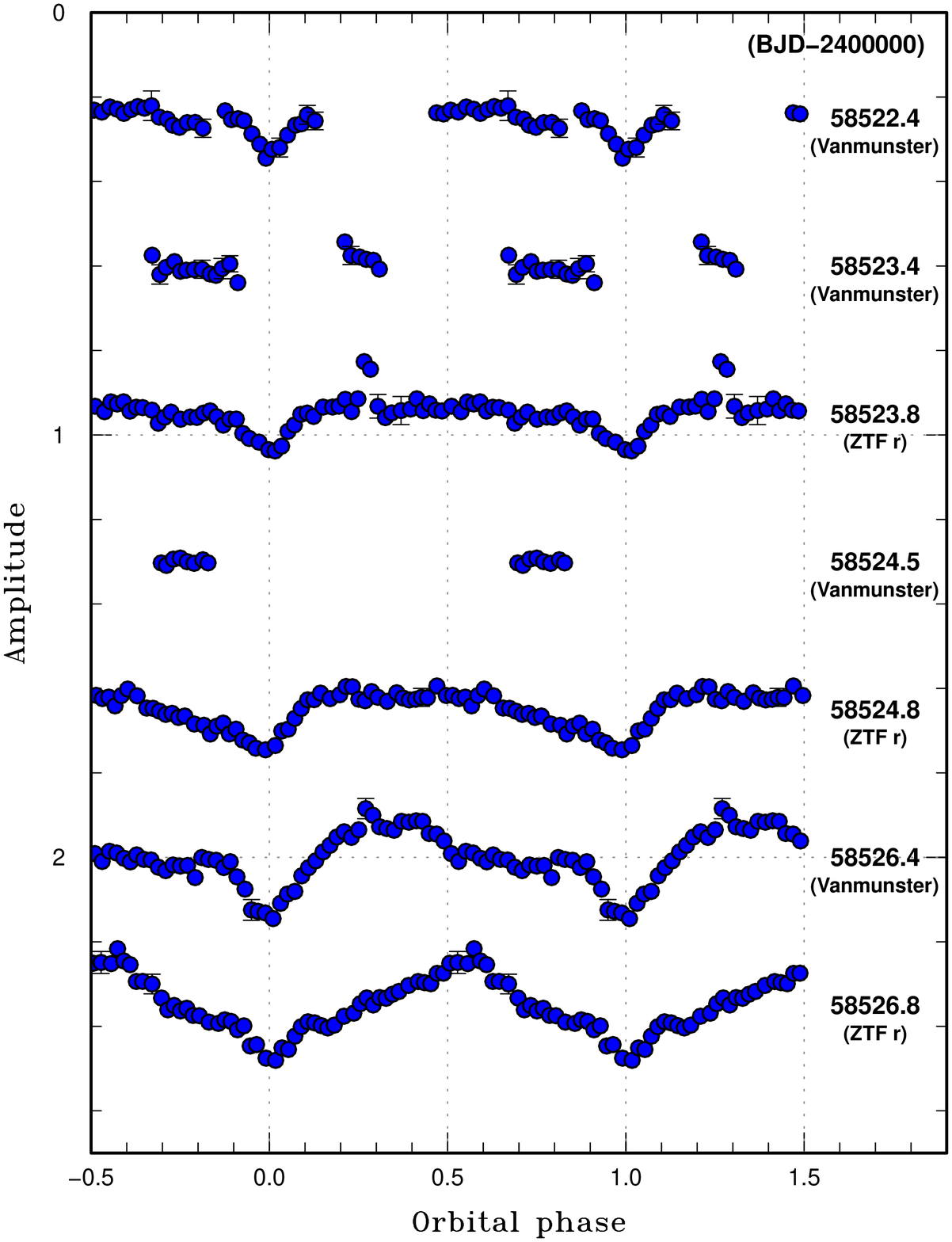}
\caption{
   Phased profiles during the 2019 February superoutburst.
   The orbital phase is defined by equation (\ref{equ:ecl}).
   The values to the right refer to the center of observational
   runs.
}
\label{fig:prof}
\end{center}
\end{figure*}

\begin{figure*}
\begin{center}
\includegraphics[width=14cm]{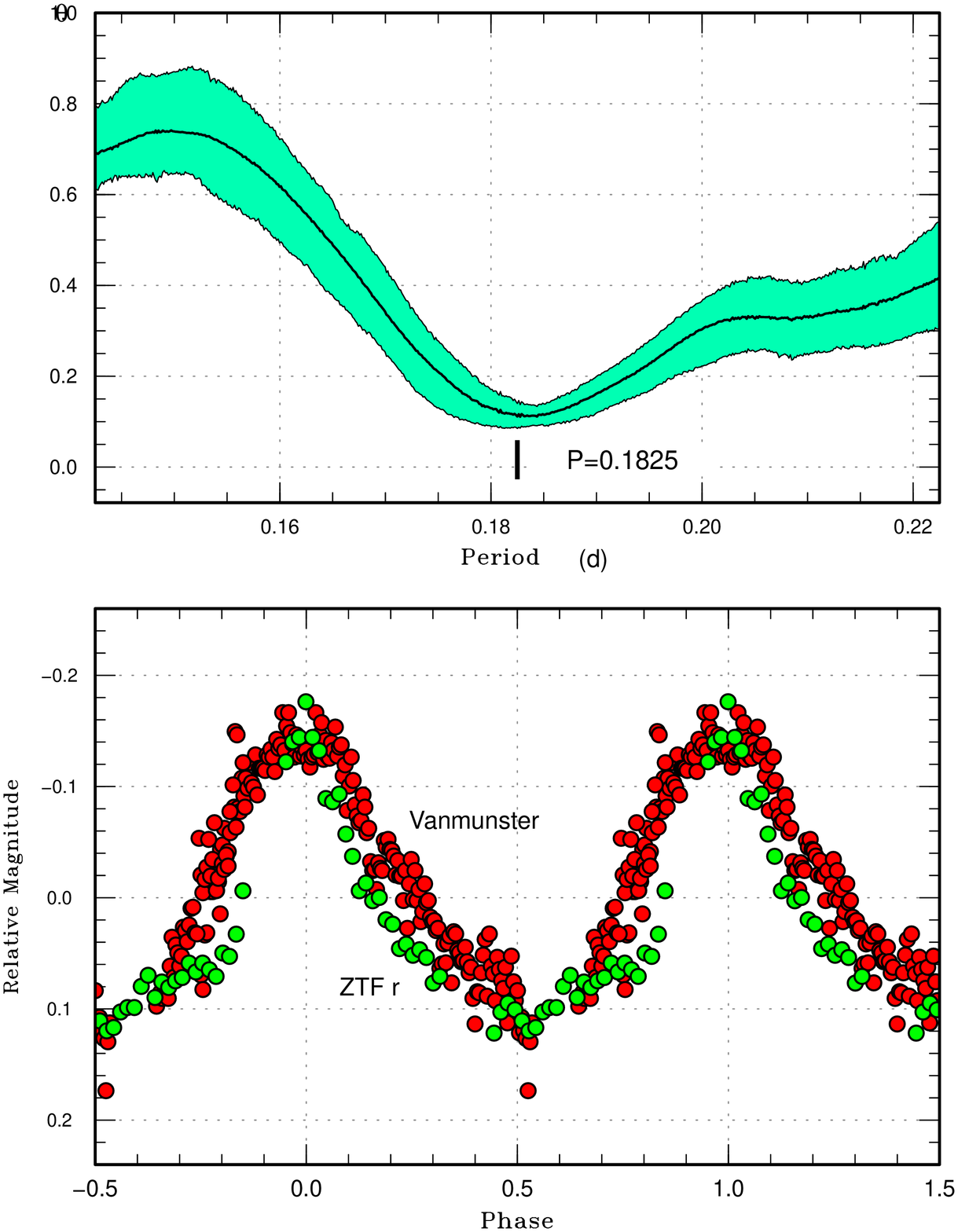}
\caption{
   Superhumps in SDSS J094002.56$+$274942.0.
   The data for BJD 2458526.33--2458526.90 were used
   after removing eclipse parts.
   (Upper): PDM analysis.
   (Lower): Phase plot.
}
\label{fig:shpdm}
\end{center}
\end{figure*}

   We show phase-averaged light curves during the 2019 February
outburst in figure \ref{fig:prof}.  The orbital phases
were obtained using equation (\ref{equ:ecl}).
The four runs (ZTF and the final night by Vanmunster) were
slightly longer than $P_{\rm orb}$
and full orbital phases were sampled in these cases.
This figure corresponds to figure 3 in \citet{kra10j0940}
(but phase-averaged).  Eclipses were present in all runs
covering the expected eclipse phase and we used these data to
determine the zero phase of equation (\ref{equ:ecl}).
The eclipses became asymmetric
during the third and fifth runs, probably reflecting
the evolution of superhumps.  In the sixth and seventh runs,
the presence of superhumps having a period longer than
$P_{\rm orb}$ is apparent.  The result of a PDM
analysis of the combined sixth and seventh runs is shown in
figure \ref{fig:shpdm} after removing the eclipse
parts (phases within $\pm$0.06).  There was some ambiguity
in the difference in the zero points between
these two runs (they were by different observers
using different equipment and filters).  We adopted
a correction of 0.050~mag to the ZTF data so that
the maxima and minima of these two runs become the same.
This ambiguity, however, did not strongly affect the result.
The resultant period was 0.1825(7)~d.   With this period,
the peaks of the superhump maxima from the two sets of
the data well agree although the waveforms were somewhat
different, probably due to the beat phenomenon between
the orbital and superhump periods (figure \ref{fig:shpdm}).
The nominal (statistical) error could be an underestimate
due to a systematic error arising from the beat phenomenon
and intrinsic variations.  A result from a short (0.57~d)
baseline would be more strongly affected by these effects
than in ordinary superhump analyses using longer baselines,
such as in \citet{Pdot}.
We, however, disregard this possibility in the following
discussions and use the nominal error.

   Although superhumps apparently grew before the sixth run,
we could not determine the period before this run
partly due to the overlapping orbital signal.
There was a gap more than 1~d before the sixth run and
it appears that the major increase of the superhump amplitude
occurred during this observational gap.

\section{Discussion}

\subsection{Mass ratio and the secondary}

   With $P_{\rm orb}$ of 0.1635015~d = 3.924~hr,
a CV on the standard evolutionary sequence is expected to
have a secondary mass of $M_2$=0.303 $M_\odot$ (main sequence)
and near-infrared absolute magnitudes of $M_J$=6.87, $M_H$=6.45
and $M_K$=6.29 \citep{kni06CVsecondary,kni07CVsecondaryerratum}.
These near-infrared absolute magnitudes can be directly comparable
to the observed 2MASS magnitudes \citep{2MASS} and the Gaia parallax
yielding a distance modulus of 9.5(2)~mag \citep{GaiaDR3}.
The observed $J$=16.15(11), $H$=15.61(16) and $K_s$=15.48(18)
correspond to the absolute magnitudes of $M_J$=6.7(2), $M_H$=6.1(3) and
$M_K$=6.0(3).
These values exclude a possibility of
an evolved donor as in other SU UMa stars with very long
$P_{\rm orb}$ [such as ASASSN-18aan ($P_{\rm orb}$=0.149454~d,
\cite{wak21asassn18aan}) and ASASSN-15cm
($P_{\rm orb}$=0.2084652~d, \cite{kat23asassn15cm})].
The secondary star in SDSS J094002.56$+$274942.0 appears
to be indistinguishable from a normal main-sequence star.
Assuming a white dwarf with an average mass $M_1$=0.82$M_\odot$
in short-period CVs \citep{sav11CVeclmass,zor11SDSSCVWDmass,
mca19DNeclipse,pal22CVparam}, the expected mass of
the secondary translates to $q$=0.37.

   What is inferred from the superhump observation?
It is well established that superhump periods vary \citep{Pdot}
and it is important which stage is used for estimating $q$
from the superhump period \citep{kat13qfromstageA,kat22stageA}.
In the case of SDSS J094002.56$+$274942.0, the superhump
observations were apparently made just around the peak
of the superhump amplitude.  It is known that stage A--B
transition [see \citet{Pdot,kat22v844her} for superhump
stages] does not match the peak of the superhump amplitude
in long-$P_{\rm orb}$ systems and stage A (judged from
the superhump period) extends slight after the maximum of
the superhump amplitude
[V1006 Cyg: \citet[][see supplementary figure]{kat16v1006cyg}
and MN Dra: \citet{Pdot6}].  We therefore consider that
the superhump period ($P_{\rm SH}$) recorded in
SDSS J094002.56$+$274942.0 reflects that of stage A superhumps.
The observed $\epsilon^* \equiv 1-P_{\rm orb}/P_{\rm SH}$
was 0.104(4).  This corresponds to $q$=0.39(3)
[see table 1 in \citet{kat22stageA}].
Considering the uncertainty in determining the superhump
period and an assumption of the white dwarf mass,
this value appears to be consistent with
a normal main-sequence secondary.
The superhump period at least does not favor a light-weight
secondary with an evolved core (as already confirmed by
the near-infrared absolute magnitudes).

\subsection{Inclination and ellipsoidal variations}\label{sec:incl}

   Eclipses observed during the superoutburst provide
an excellent opportunity for modeling the binary since
the disk radius is expected to be the radius of
the 3:1 resonance.
Assuming an optically thick standard disk, inclinations
needed to reproduce the observed eclipse depth (0.11~mag)
are given in table \ref{tab:incl}.  We used three $q$
values (0.37 from the standard evolutionary sequence
and an average-mass white dwarf, 0.39 from stage A
superhumps and 0.42 as the upper limit).
The uncertainties of inclinations were 0.5$^\circ$ and
the values were rounded to 0.5$^\circ$.

\begin{figure*}
\begin{center}
\includegraphics[width=16cm]{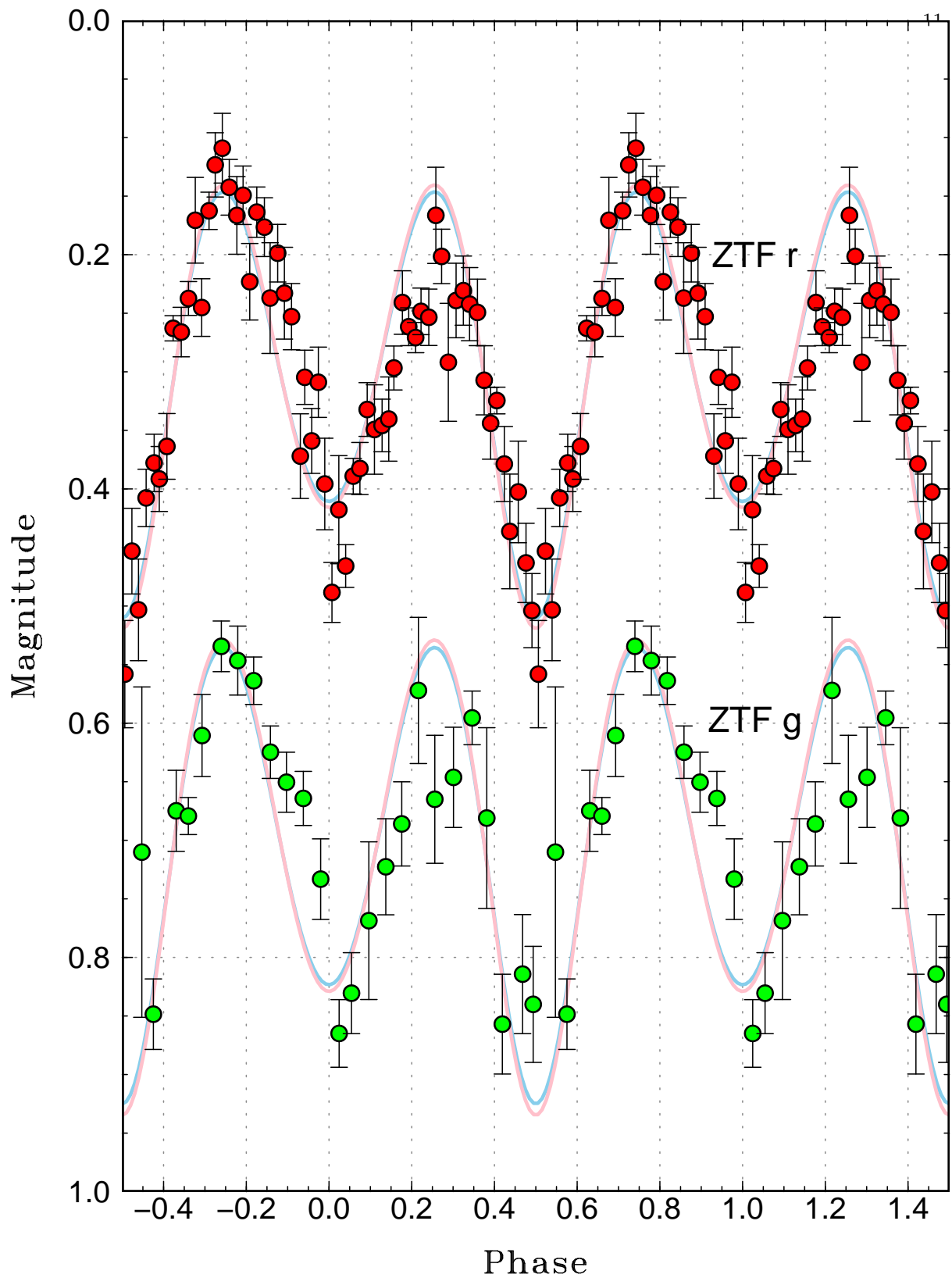}
\caption{
   Orbital profile of SDSS J094002.56$+$274942.0 in quiescence using
   ZTF $r$ and $g$ data.  The orbital phase is defined by equation
   (\ref{equ:ecl}).  The plots were shifted by 0.4~mag
   between different bands.
   The solid curves represent ellipsoidal variations
   expected for $q$=0.43, $i$=70.0$^\circ$ (skyblue)
   and $q$=0.37, $i$=71.0$^\circ$ (pink) (see text).
}
\label{fig:quilc}
\end{center}
\end{figure*}

   We followed the case of ASASSN-15cm \citep{kat23asassn15cm}
in modeling the quiescent light curve.
Using Ellipsoidal Modulation Light Curve Generator by
M.~Uemura (2006), which was based on
\citet{oro97j1655,oro97j1655erratum}, we could reproduce
ellipsoidal variations ZTF $r$ and $g$ observations
in figure \ref{fig:quilc}.  We used $T_{\rm eff}$=3500~K,
$\log g$=4.727, assuming a main-sequence secondary in
\citet{kni06CVsecondary,kni07CVsecondaryerratum} and
gravity darkening and limb-darkening coefficients given
in \citet{cla11limbdark} (solar metallicity and a microturbulent
velocity of 2~km s$^{-1}$ as typical values were used).
As in the case of ASASSN-15cm, outbursts in
SDSS J094002.56$+$274942.0 were rare and the mass-transfer rate
is expected to be low.  The contribution from the disk appears
to be negligible in quiescence.  Both $q$=0.37 and $q$=0.42
cases sufficiently fit the observations with the inclinations
determined by eclipse analysis.  It turned out that quiescent
ellipsoidal variations could not constrain the binary parameter
better than what we obtained from superhump and eclipse analysis.
This result, however, confirmed that our basic picture
(unevolved main-sequence secondary) is correct.

\begin{table*}
\caption{Eclipse modeling during superoutburst.}
\label{tab:incl}
\begin{center}
\begin{tabular}{cc}
\hline
$q$ & Inclination ($^\circ$) \\
\hline
0.37 & 71.0 \\
0.39 & 70.5 \\
0.42 & 70.0 \\
\hline
\end{tabular}
\end{center}
\end{table*}

\subsection{Implications of the present discovery}

   As stated in section \ref{sec:intro}, it has long been
considered that CVs above the period gap showing superhumps
are almost entirely novalike variables with a thermally stable
(high-state) disk.  It has been considered that superhumps can
grow only in such systems in sufficiently long high states
under a weaker resonance effect.  Indeed, some established
SU UMa-type dwarf novae above the period gap have a secondary
with an evolved core such as in OT J002656.6$+$284933
(=CSS101212:002657$+$284933, $P_{\rm SH}$=0.13225~d: \cite{kat17j0026}),
ASASSN-18aan ($P_{\rm orb}$=0.149454~d: \cite{wak21asassn18aan})
and ASASSN-15cm ($P_{\rm orb}$=0.208466~d: \cite{kat23asassn15cm}).
\citet{wak21asassn18aan} suggested that the 3:1 resonance
in long-$P_{\rm orb}$ dwarf novae is more difficult
to develop around the stability border
than in the case of the 2:1 resonance.
The present discovery of an SU UMa-type dwarf nova with
$P_{\rm orb}$=0.1635015~d with a normal main-sequence secondary
is against this interpretation.  Even with $q$=0.39(3),
this dwarf nova apparently started developing superhumps
soon after it reached the plateau phase and fully developed
superhumps were recorded within 6~d of the start of
the plateau phase.  This behavior is common to other
SU UMa-type dwarf novae and there was no significant difference
in the development of superhumps.
This case supports the idea that the 3:1 resonance triggered
a superoutburst even in this extreme case.

   The superoutburst, however, terminated rather quickly and was
followed by a rebrightening.  These features suggest that
the hot state was difficult to maintain by the 3:1 resonance
in an extreme $q$ case --- a mechanism proposed in V1006 Cyg
and CS Ind which simulated a WZ Sge-type phenomenon
\citep{kat16v1006cyg,kat19csind} ---
and that the matter left in the disk due to the early quenching
of the superoutburst resulted a rebrightening, which is
usually a feature in short-$P_{\rm orb}$ systems
\citep{kat98super,kat15wzsge}.

   Continued monitoring of this object is required, and
better determination of the superhump period and its evolution
will clarify what is happening in long-$P_{\rm orb}$
and high-$q$ SU UMa-type dwarf novae.

\section*{Acknowledgements}

This work was supported by JSPS KAKENHI Grant Number 21K03616.
The authors are grateful to the ZTF and ATLAS teams
for making their data available to the public.
We are grateful to VSOLJ observers (particularly
Yutaka Maeda) who reported snapshot CCD photometry of
SDSS J094002.56$+$274942.0.
We are also grateful to Makoto Uemura for sharing the code of
Ellipsoidal Modulation Light Curve Generator,
Naoto Kojiguchi for helping downloading
the ZTF and TESS data and Yasuyuki Wakamatsu for
converting the data reported to the VSNET Collaboration.

Based on observations obtained with the Samuel Oschin 48-inch
Telescope at the Palomar Observatory as part of
the Zwicky Transient Facility project. ZTF is supported by
the National Science Foundation under Grant No. AST-1440341
and a collaboration including Caltech, IPAC, 
the Weizmann Institute for Science, the Oskar Klein Center
at Stockholm University, the University of Maryland,
the University of Washington, Deutsches Elektronen-Synchrotron
and Humboldt University, Los Alamos National Laboratories, 
the TANGO Consortium of Taiwan, the University of 
Wisconsin at Milwaukee, and Lawrence Berkeley National Laboratories.
Operations are conducted by COO, IPAC, and UW.

The ztfquery code was funded by the European Research Council
(ERC) under the European Union's Horizon 2020 research and 
innovation programme (grant agreement n$^{\circ}$759194
-- USNAC, PI: Rigault).

This work has made use of data from the Asteroid Terrestrial-impact
Last Alert System (ATLAS) project.
The ATLAS project is primarily funded to search for
near earth asteroids through NASA grants NN12AR55G, 80NSSC18K0284,
and 80NSSC18K1575; byproducts of the NEO search include images and
catalogs from the survey area. This work was partially funded by
Kepler/K2 grant J1944/80NSSC19K0112 and HST GO-15889, and STFC
grants ST/T000198/1 and ST/S006109/1. The ATLAS science products
have been made possible through the contributions of the University
of Hawaii Institute for Astronomy, the Queen's University Belfast, 
the Space Telescope Science Institute, the South African Astronomical
Observatory, and The Millennium Institute of Astrophysics (MAS), Chile.

\section*{List of objects in this paper}
\xxinput{objlist.inc}

\section*{References}

We provide two forms of the references section (for ADS
and as published) so that the references can be easily
incorporated into ADS.

\newcommand{\noop}[1]{}\newcommand{\hyphalt}{-}

\renewcommand\refname{\textbf{References (for ADS)}}

\xxinput{j0940aph.bbl}

\renewcommand\refname{\textbf{References (as published)}}

\xxinput{j0940.bbl.vsolj}

\end{document}